\documentclass{article} 
\usepackage[final]{colm2026_conference}

\usepackage{hyperref}
\usepackage{url}
\usepackage{booktabs}

\usepackage{amsmath}
\usepackage{cleveref}
\usepackage{xspace}
\usepackage{array}
\usepackage{balance}
\usepackage{graphicx}
\usepackage{microtype}
\usepackage{url}
\usepackage{changepage}

\newcommand{\bench}{\textsc{UrbanContrastiveQA}\xspace}
\newcommand{\wrongscale}{\emph{wrong-scale comparison}\xspace}

\newcommand{\toolfmt}[1]{%
  \begingroup
  \setlength{\fboxsep}{1.2pt}%
  \colorbox{gray!12}{\textbf{\textsf{\footnotesize #1}}}%
  \endgroup\xspace
}

\newcommand{\fCountOnly}{\toolfmt{Count-only}}
\newcommand{\fPlusBaseline}{\toolfmt{+Baseline}}
\newcommand{\fAskScore}{\toolfmt{Ask-score}}
\newcommand{\fPlusServerScore}{\toolfmt{+Server-score}}
\newcommand{\fFullView}{\toolfmt{Full-view}}
\newcommand{\fScoreGap}{\toolfmt{Med. $|\Delta z|$}}

\usepackage{lineno}

\definecolor{darkblue}{rgb}{0, 0, 0.5}
\hypersetup{colorlinks=true, citecolor=darkblue, linkcolor=darkblue, urlcolor=darkblue}

\title{Measuring Decision-Scale Use in Tool-Augmented LLMs: \\A Contrastive Urban Benchmark}

\author{Ray Chen\\
Department of Computer \& Information Science Engineering\\
University of Florida\\
\texttt{chenz1@ufl.edu}
\AND
Vivian Wong\\
College of Design, Construction and Planning\\
University of Florida\\
\texttt{vivian.wong@ufl.edu}
\AND
Christan Grant\\
Department of Computer \& Information Science Engineering\\
University of Florida\\
\texttt{christan@ufl.edu}
}

\begin{document}

\ifcolmsubmission
\linenumbers
\fi

\maketitle

\begin{abstract}
Urban decision-support often asks whether activity is unusually high or low for a specific place, not which place has the larger raw count. Twenty pickups in a quiet neighborhood can be more abnormal than 180 at an airport. We introduce \bench, a benchmark that asks whether tool-augmented language models can make this baseline-relative comparison. Each item pairs two urban situations from public mobility data in NYC, Chicago, and Seattle, labeled by how far current activity deviates from that place's historical baseline. We evaluate six instruction-tuned models under five tool-output formats. With only raw counts, models often pick the larger number even when it is less abnormal for its zone. Server-computed baseline scores and ordinal labels raise accuracy, but gains vary by model. For heterogeneous urban feeds, tool interfaces need to expose local baselines, not just activity volumes. We release the pair bank, labels, scoring scripts, and data card.

\end{abstract}

\section{Introduction}
\label{sec:intro}

Urban decision-support often asks whether activity is unusually high or low for a \emph{specific}
place, not whether one place has a larger raw count than another.
For example, 180 airport pickups can be \emph{below} normal for that zone while 20 pickups in a
quiet residential area can be \emph{above} normal, because each place has its own historical
baseline.
Comparing raw counts across heterogeneous zones leads to \wrongscale{}: choosing based on raw
magnitude when the task calls for deviation from a local baseline.

Tool-augmented language models read structured numbers from retrieval and tool APIs
\citep{lewis2020retrieval, yao2022react}; recent work has moved to city-scale and geospatial
settings \citep{manvi2024geollm, li2024urbangpt, li2025urban}.
Retrieval evaluations usually ask whether the right evidence was retrieved or whether irrelevant
context misleads the model \citep{shi2023distracted}.
General-purpose LM benchmarks \citep{hendrycks2021measuring} and numerical table QA tasks rarely
check whether a model compares urban activity to each place's own local baseline when tool
outputs supply counts on different scales.

We introduce \bench{}, a benchmark for baseline-relative urban comparison in tool-augmented LLMs.
Each item presents two urban situations drawn from public mobility data in New York City, Chicago,
and Seattle.
Both sides include location, time, and weather context, along with a historical baseline for that
place.
The model must choose which situation is more above normal relative to its own baseline, not
which side has the larger raw count.

\textbf{Our contributions are:}
(1)~We formulate baseline-relative urban comparison as a tool-augmented LLM evaluation task.
(2)~We construct \bench{}, a benchmark of 432 paired urban situations from NYC, Chicago, and
Seattle, with labels based on per-zone deviations from historical baselines.
(3)~We evaluate six instruction-tuned models across multiple tool-output formats and show that
\fCountOnly often leads models toward raw-magnitude choices.
(4)~We run swap controls to see whether models use attached baseline fields or fall back on place
names and raw magnitudes.
(5)~We release the benchmark, labels, scoring scripts, and data card.

\begin{figure}[!t]
  \centering
  \includegraphics[width=0.8\textwidth, height=0.5\textwidth]{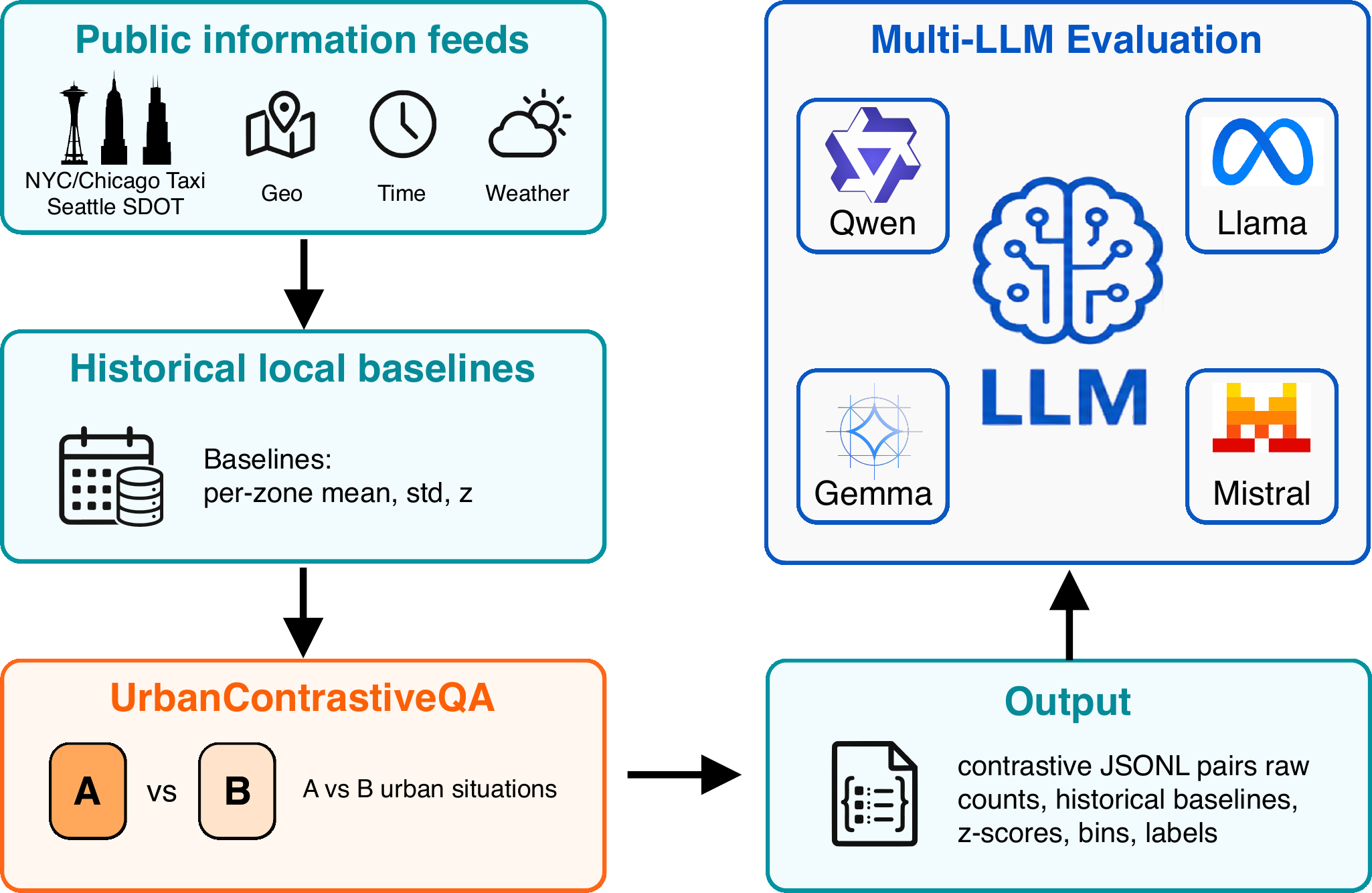}\\[2pt]
  {\scriptsize Tool-output formats: \fCountOnly$|$\fPlusBaseline$|$\fAskScore$|$\fPlusServerScore$|$\fFullView}
\caption{Overview of \bench. Public mobility feeds are converted into local baselines, paired comparison items, and fixed tool calls shown under five output formats. The benchmark tests whether models compare local deviations rather than raw counts.}
  \label{fig:bench_overview}
\end{figure}

\section{Benchmark Design}
\label{sec:benchmark}

The pipeline starts from public mobility feeds, builds per-zone baseline statistics, forms
paired items, renders tool output in five formats, and evaluates models, as shown in
Figure~\ref{fig:bench_overview}.

Each item asks the model to compare two urban situations $\langle p_A, p_B \rangle$ from the
same corpus.
Some pairs are easy because the larger raw count is also more abnormal; others are hard because
the larger raw count is less abnormal under the baseline-relative label.
We call these \emph{aligned} and \emph{conflict} pairs.
Given each side's location, time, and weather, the model answers one baseline-relative
question: \emph{``Which situation is more above normal relative to its own historical baseline?
Answer A or B.''}
We score abnormality as the observed count minus that place's historical average, divided by its
historical variation.
For a region $r$, month $m$, and hour-of-week $h$, the score is
\[
z = \frac{x-\mu_{r,m,h}}{\sigma_{r,m,h}}
\]
where $x$ is the observed activity count, $\mu_{r,m,h}$ is the historical average, and $\sigma_{r,m,h}$ is the historical standard deviation.

We compare five ways of showing the same tool result.
\fCountOnly returns only the observed activity count.
\fPlusBaseline adds the local historical average and variation.
\fAskScore gives the model the count and baseline, then asks it to compute how unusual the count is.
\fPlusServerScore returns a server-computed baseline-relative score and an ordinal label such as below, near, or above normal.
\fFullView returns all baseline-related fields, including the count, baseline scope, historical average, historical variation, baseline-relative score, ordinal label, and context.

\subsection{Related Work}
\label{sec:related}

Tool-augmented LMs learn to call external APIs from demonstrations and reasoning traces
\citep{yao2022react, lewis2020retrieval}.
Most faithfulness work asks whether models retrieve the right passage or ignore irrelevant
context \citep{shi2023distracted}.
Agent benchmarks score end-to-end tool execution \citep{liu2023agentbench}.
We fix the tool call and vary only how the same counts are rendered, so failures trace to
\wrongscale{} rather than API selection or parsing.
Numerical QA tasks often test arithmetic over fixed tables; general LM suites
\citep{hendrycks2021measuring} and urban LLM work
\citep{manvi2024geollm, li2024urbangpt, li2025urban} rarely ask whether a retrieved value is
above \emph{its own} local baseline when two zones sit on different raw-count scales.
\bench{} supplies pairwise probes with ground-truth per-zone labels and conflict strata aimed at
that gap.

\subsection{Data Sources and Baseline Construction}
\label{sec:construction}

Baseline-relative ground truth comes from a reusable pipeline over public feeds.
Data sources include NYC TLC Yellow Taxi trip records, Chicago Transportation Network Provider trip records aggregated by community area, Seattle SDOT bicycle/pedestrian counter feeds, and NOAA ISD-Lite weather observations~\cite{nyctlc_trip_records,chicago_tnp_trips,seattle_open_data,noaa_isd}.
Each stream is harmonized to hourly counts indexed by
$(z,\text{year},\text{month},\text{hour\_of\_week})$.

\emph{Baseline formula:} per-zone baselines use 2019 and 2022, excluding pandemic years, with
held-out evaluation on 2023:
$\mu_{z,m,h}=\mathrm{mean}(\text{count})$,
$\sigma_{z,m,h}=\max(\mathrm{std}(\text{count}),1.0)$.
Ground-truth $z$ and ordinal labels, using $|z|>0.5$, are stored per side; aligned/conflict labels
follow a stereotype heuristic versus the $z$ ordering.

\emph{Label reliability:} a manual audit of 20 weather-conflict pairs found that the higher score
side matched the automated winner in all audited cases.
On the full pair bank, single-year baselines agree with the ground-truth score winner on
$\ge 91\%$ of evaluable pairs. Median/MAD and percentile-rank relabelings flip many winners,
with agreement at $\le 54\%$; diagnostics are included in the artifact bundle.
The pipeline uses only aggregated hourly open-data volumes and no rider-level PII.
The released data card documents further construction details, including the $\sigma$ fallback
hierarchy, near-boundary flags, stable-subset evaluation, and the scarcity of spatial-conflict
pairs in the main bank ($n{=}26$).

Because abnormality is a measurement construct rather than a directly observed fact, these relabeling disagreements are informative rather than incidental. We use z-score labels as the primary operational definition and include stability diagnostics so users can distinguish interface failures from construct-choice sensitivity.

\subsection{Pair Construction}

The full \emph{pair bank} contains 432 unordered pairs, or $864$ orientations, with chance at
$0.5$. It is stratified by four \emph{types}: temporal, spatial, weather, and cross. We also track
aligned/conflict status; see Table~\ref{tab:bench_stats}.
Temporal pairs contrast hour-of-week and month at fixed geography; spatial pairs contrast two
zones in the same city at matched month and hour-of-week; weather pairs contrast weather bucket
at fixed $(\mathrm{city},h,m)$; cross pairs link situations across cities.
Each JSONL row nests full probe records for sides A and B with per-side count, baseline mean
and standard deviation, baseline-relative score, ordinal label, geography, time, and weather.
One Chicago spatial conflict, for example, has O'Hare's higher raw count below normal relative to
its own baseline while a quiet zone is above normal.
We also release a \emph{diagnostic subset} of 200 pairs, or $400$ orientations, where raw-count
comparison is intentionally misleading.

\begin{table}[t]
\centering\normalsize
\setlength{\tabcolsep}{3.5pt}
\begin{tabular}{@{}lrrrr@{}}
\toprule
\textbf{Type} & \textbf{Pairs} & \textbf{Aligned} & \textbf{Conflict} & \fScoreGap \\
\midrule
\multicolumn{5}{@{}l}{\emph{Pair bank (NYC, Chicago, Seattle)}}\\
Temporal  & 200 & 141 & 59  & 0.71 \\
Spatial   &  41 &  15 & 26  & 1.02 \\
Weather   &  91 &  45 & 46  & 0.65 \\
Cross     & 100 & 100 &  0  & --   \\
\textbf{Total} & \textbf{432} & \textbf{301} & \textbf{131} & 0.74 \\
\midrule
\multicolumn{5}{@{}l}{\emph{Diagnostic subset ($n{=}200$ pairs)}}\\
\textbf{Subset} & \textbf{200} & 0 & \textbf{200} & -- \\
\bottomrule
\end{tabular}
\caption{\bench statistics.
Conflict pairs are where stereotype or raw-count following disagrees with the baseline-relative score label;
spatial conflict is the smallest slice ($26$ pairs) but the hardest in our runs. \fScoreGap is the absolute difference between the two sides' baseline-relative scores.
}
\label{tab:bench_stats}
\end{table}

\textbf{Pairwise prompt for all items.}
\begin{quote}
\begin{adjustwidth}{-5pt}{-5pt}
Which situation is more above normal relative to its own historical baseline?\\
That is, which has the higher baseline relative z-score?\\
Situation A: urban activity in [location A] on [time A] during [weather A].\\
Situation B: urban activity in [location B] on [time B] during [weather B].\\
Answer with exactly: A or B.
\end{adjustwidth}
\end{quote}

Each item stores the unordered pair identifier, ordered A/B presentation, source city, zone
identifiers, time fields, weather bucket, raw count, baseline statistics, computed baseline-relative score,
ordinal label, aligned/conflict tag, and ground-truth winner.
Evaluation uses the ordered presentation; pair-level accuracy requires both orientations correct.
Invalid and non-A/B outputs count as incorrect.

\subsection{Tool-Output Rendering Conditions}
\label{sec:rendering}
We evaluate six open instruction-tuned models at temperature~0: Qwen3-30B-A3B and Qwen3-4B
\citep{yang2025qwen3}, Llama-4-Scout-17B \citep{llama4scout2025}, Gemma-3-27B-it and Gemma-3-4B-it
\citep{gemma2025}, and Mistral-Small-3.1-24B \citep{mistralsmall31_2025} under vLLM
with max length 4096. The runs used $\sim$60 GPU-hours in total.

We evaluate the five tool-output formats introduced above under identical tool calls. A deterministic \emph{program $z$-rule} comparing stored $z_A$ vs.\ $z_B$ provides a $1.000$ upper bound.

\textbf{Tool payloads for a schematic conflict example.}
\fCountOnly attaches incomparable counts; \fFullView attaches the per-side decision-scale fields used by the label, and a raw-magnitude policy picks B, but the baseline-relative label is A.
{\sloppy\begin{quote}
\begin{adjustwidth}{-10pt}{-10pt}
{[}Tool A: \{"zone\_avg\_pickups": 20.0, "scope": "zone"\}{]}\\
{[}Tool B: \{"zone\_avg\_pickups": 180.0, "scope": "zone"\}{]}\\
{[}Tool A: baseline\_mean=8, baseline\_std=6, z\_score=2.00, bin=above normal{]}\\
{[}Tool B: baseline\_mean=250, baseline\_std=50, z\_score=-1.40, bin=below normal{]}
\end{adjustwidth}

\end{quote}\fussy}

\subsection{Evaluation and Controls}
\label{sec:baselines}

We report orientation-level accuracy for each ordered presentation and pair-level accuracy, which
requires both orientations to be correct. Uncertainty uses pair-clustered bootstrap over the 432
unordered pairs.
Two validity controls test whether models follow side-attached evidence:
\emph{raw-number swap} exchanges the two raw payloads while keeping the question fixed;
\emph{baseline-field swap} attaches the opposite side's baseline score and ordinal label fields.
Simple non-LM rules bracket model behavior: a raw-magnitude heuristic that picks the larger
zone-average pickup count reaches $0.383$ overall and $1.000/0.000$ on spatial aligned/conflict slices; a
local-score heuristic is $1.000$ everywhere. Conflict strata therefore separate raw following
from baseline-relative truth.
Pairs, ground-truth scores, scoring scripts, and metric definitions will be publicly released under
an open license upon acceptance.


\begin{table*}[t]
\centering
\small
\setlength{\tabcolsep}{2pt}
\resizebox{\textwidth}{!}{%
\begin{tabular}{@{}lccccc@{}}
\toprule
\textbf{Model}
& \fCountOnly
& \fPlusBaseline
& \fAskScore
& \fPlusServerScore
& \fFullView \\
\midrule
Qwen3-4B          & 0.396 [0.351, 0.439] & 0.728 [0.690, 0.764] & 0.678 [0.644, 0.712]
  & \textbf{0.852 [0.822, 0.880]} & 0.809 [0.779, 0.837] \\
Qwen3-30B-A3B     & 0.312 [0.275, 0.351] & 0.743 [0.713, 0.772] & 0.722 [0.686, 0.757]
  & 0.881 [0.858, 0.903] & \textbf{0.934 [0.917, 0.950]} \\
Gemma-3-4B        & 0.481 [0.448, 0.515] & 0.725 [0.688, 0.760] & 0.742 [0.705, 0.777]
  & \textbf{0.769 [0.738, 0.797]} & 0.696 [0.656, 0.733] \\
Gemma-3-27B       & 0.394 [0.351, 0.436] & 0.736 [0.701, 0.769] & 0.245$^\dagger$ [0.206, 0.286]
  & \textbf{0.841 [0.810, 0.870]} & 0.759 [0.720, 0.796] \\
Llama-4-Scout-17B & 0.251 [0.218, 0.287] & 0.693 [0.653, 0.733] & 0.010$^\dagger$ [0.001, 0.020]
  & 0.909 [0.885, 0.931] & \textbf{0.932 [0.913, 0.949]} \\
Mistral-Small-24B & 0.383 [0.339, 0.427] & 0.762 [0.725, 0.796] & 0.728 [0.689, 0.766]
  & \textbf{0.932 [0.912, 0.949]} & 0.828 [0.793, 0.859] \\
\bottomrule
\end{tabular}%
}
\caption{Reference results on the pair bank. \fCountOnly stays near the raw-count heuristic, while \fPlusServerScore and \fFullView improve all six models. \fPlusServerScore is best for four of six models, while \fFullView is best for Qwen3-30B and Llama. Values are orientation accuracy with pair-clustered 95\% CIs over 432 unordered pairs. \textbf{Bold} marks the best format in each model row. $^\dagger$Under \fAskScore{}, many responses fail to parse as A or B. Llama-4-Scout parses $9/864$ orientations, and Gemma-3-27B parses a partial set.}
\label{tab:format_ladder}
\end{table*}

\section{Results}
\label{sec:findings}

\begin{figure}[t]
  \centering
  \includegraphics[width=0.8\textwidth, height=0.6\textwidth]{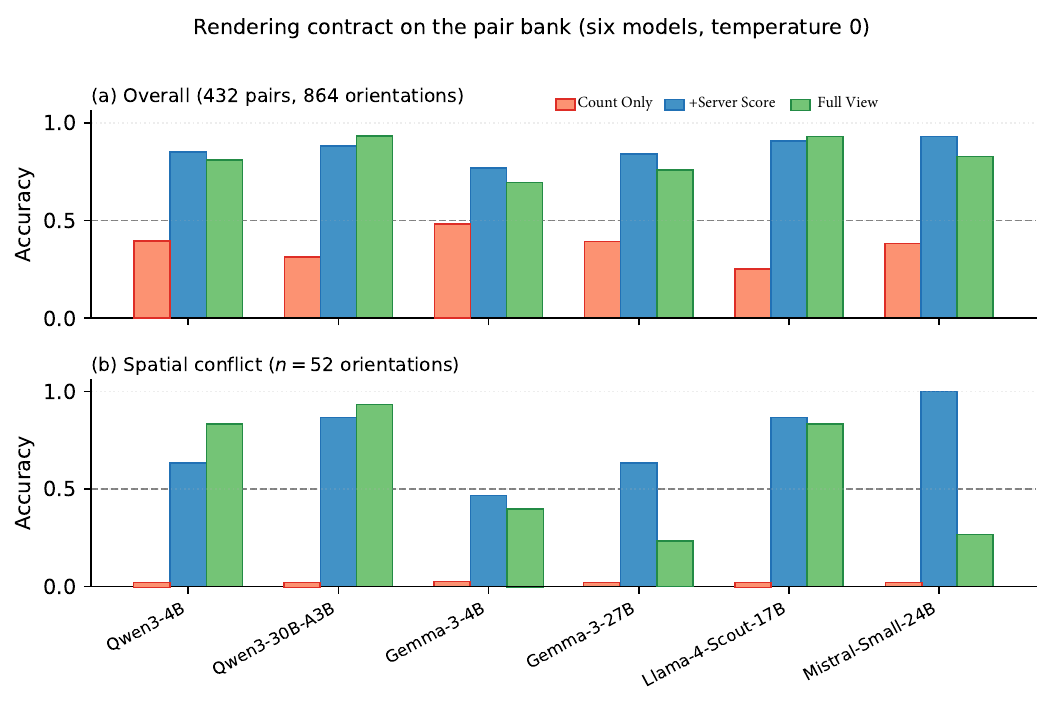}
\caption{Tool-output format changes model accuracy on the pair bank.
(a) \fCountOnly stays near chance overall, while \fPlusServerScore and \fFullView improve all six models.
(b) On spatial-conflict pairs, \fCountOnly gives nearly 0.000 accuracy for every model, showing that raw counts do not support local-baseline comparison.}
  \label{fig:main_story}
\end{figure}

The spatial problem shows up when raw magnitude and local abnormality diverge.
With \fCountOnly, all six models stay near the raw-count heuristic on the full pair bank, with
accuracies from 0.25 to 0.48, and score 0.000 on the spatial-conflict slice shown in
Figure~\ref{fig:main_story}b.
That slice is small, with $26$ pairs, but telling: month and hour are fixed while zones with
different local baselines are compared, so the model must judge deviations from local norms, not
absolute volume.
For example, on Qwen3-4B, aligned spatial pairs reach $0.981$ across $n{=}30$ orientations, while
spatial-conflict pairs drop to $0.000$ across $n{=}52$.
Cross-city pairs preserve realistic variety, with $n{=}100$ and no conflicts in that stratum, but
they do not trap \wrongscale{}; pooled accuracy alone hides the spatial scale problem.

Local baseline information changes behavior across the pair bank.
\fPlusServerScore and \fFullView improve every model, though the winning format depends on the
model; see Figure~\ref{fig:main_story}a and Table~\ref{tab:format_ladder}.
\fPlusServerScore wins for four of six models; \fFullView wins for Qwen3-30B and Llama.
Urban tool interfaces should return the observed count together with local baseline context, not
raw volumes alone.
On Qwen3-4B, weather-conflict accuracy rises from $0.305$ with \fCountOnly to $0.951$ with
\fPlusServerScore, while \fPlusBaseline and \fAskScore stay less reliable.

The diagnostic subset stress-tests the spatial case: every item is built so raw-count comparison
disagrees with the local-baseline label. Table~\ref{tab:stress_subset} reports these results.
On this subset, Mistral-Small reaches $0.986$ [0.972, 0.987] with \fPlusServerScore, and
\fFullView reaches $0.974$ [0.963, 0.980], while the program score oracle is $1.000$.
Gemma-3-4B \fFullView at $0.420$ is below the $0.5$ chance rate, consistent with reverse raw-magnitude
following when every pair is conflict-only.
Raw-number swap is omitted from the table because it is $1.000$ on this subset by construction.

\begin{table}[t]
\centering\normalsize
\setlength{\tabcolsep}{3pt}
\begin{tabular}{@{}lcc@{}}
\toprule
\textbf{Model} & \fFullView & \fPlusServerScore \\
\midrule
Qwen3-4B          & \textbf{0.787 [0.745, 0.830]} & 0.608 [0.552, 0.662] \\
Qwen3-30B-A3B     & \textbf{0.917 [0.890, 0.943]} & 0.833 [0.792, 0.870] \\
Gemma-3-4B        & 0.420 [0.372, 0.468] & \textbf{0.422 [0.372, 0.472]} \\
Gemma-3-27B       & 0.573 [0.522, 0.623] & \textbf{0.620 [0.565, 0.675]} \\
Llama-4-Scout-17B & 0.772 [0.733, 0.812] & \textbf{0.835 [0.797, 0.873]} \\
Mistral-Small-24B & 0.974 [0.963, 0.980] & \textbf{0.986 [0.972, 0.987]} \\
\bottomrule
\end{tabular}
\caption{Diagnostic subset where raw-count comparison disagrees with the local-baseline label.
Server-computed scores help, but several models stay below the program oracle.
\textbf{Bold} marks the better format in each model row.}
\label{tab:stress_subset}
\end{table}

Swap controls check whether models follow side-attached evidence or fall back on place priors.
Raw-number swap moves Qwen3-4B accuracy from $0.396$ to $0.597$ and flips the spatial-aligned
pattern from $0.981$ to $0.019$; Baseline-field swap reduces Qwen3-4B accuracy to 0.434 under +Server-score and 0.569 under \fFullView.
Models do respond to attached fields, but the rendering contract still matters.
On conflict orientations, Qwen3-4B stays confident while wrong. Mean confidence is $9.21$ with
\fCountOnly, $9.79$ with \fPlusServerScore, and $9.53$ with \fFullView.
When \fFullView trails \fPlusServerScore, the extra normalized fields may drown out the
per-side baseline-relative score and ordinal label that the leaner format keeps in view.
A program rule comparing stored $z_A$ and $z_B$ reaches $1.000$ on all $864$ orientations;
\fPlusServerScore and \fFullView test whether the interface exposes decision-scale statistics, not whether
the LM computed $z$ from \fCountOnly.

For heterogeneous urban feeds, the tool interface should pair each observed count with local
baseline context: scope such as zone, borough, or city; historical mean and standard deviation;
server-computed baseline-relative score; ordinal label such as below, near, or above normal; and
side-local geography, time, and weather.
\fFullView adds more baseline-normalized fields; \fPlusServerScore keeps the raw count visible
while still exposing the decision scale used by the label.
\bench{} targets baseline-relative comparison over public activity feeds, not general urban
forecasting.
The main pair bank keeps naturally occurring temporal, spatial, weather, and cross-city
comparisons; the diagnostic subset holds cases where raw magnitude and local abnormality
disagree.

\section{Conclusion}
\label{sec:conclusion}

\bench{} asks whether tool-augmented LLMs compare urban activity to local baselines instead of
raw counts.
It pairs items from three US cities with conflict strata, a diagnostic subset, and swap controls.
\fCountOnly and \fPlusBaseline fail to support baseline-relative comparison reliably;
\fPlusServerScore helps most models the most.
Pooled accuracy alone is a weak signal for heterogeneous urban feeds; conflict strata and swap
controls catch \wrongscale{} failures that averages miss.
More broadly, the benchmark shows that tool-augmented evaluation should specify not only the evidence source but also the decision scale on which model answers are judged.

\balance
\bibliographystyle{colm2026_conference}
\bibliography{reference}

\end{document}